\pdfoutput=1

\documentclass[11pt]{article}

\usepackage[preprint]{acl}

\usepackage{times}
\usepackage{latexsym}
\usepackage{amssymb}

\usepackage[T1]{fontenc}

\usepackage[utf8]{inputenc}

\usepackage{microtype}

\usepackage{inconsolata}
\usepackage{graphicx}
\usepackage{booktabs}
\usepackage{subcaption}
\usepackage{array}

\usepackage{amsmath}

\usepackage{tcolorbox}
\usepackage{fvextra}

\usepackage{comment}
\usepackage{xcolor, pifont}

\usepackage{booktabs, multirow, makecell, rotating, xcolor, colortbl}
\usepackage[flushleft]{threeparttable}

\newcommand{\rot}[1]{\rotatebox{90}{#1}}
\definecolor{grpA}{HTML}{E8F4FF}     
\definecolor{grpB}{HTML}{FFE8EB}     
\definecolor{grpC}{HTML}{E8FFEF}     
\definecolor{grpD}{HTML}{FFF0E0} 
\definecolor{grpE}{HTML}{FFF0F6} 
\definecolor{avgcol}{gray}{0.9}      
\definecolor{dscol}{gray}{0.95}      

\usepackage{siunitx}
\newif\ifptitle
\newif\ifpnumber
\newcounter{para}

\title{Context is Gold to find the Gold Passage: Evaluating and Training Contextual Document Embeddings}

\author{Max Conti$^*$$^{1,4}$ \quad Manuel Faysse\thanks{Equal Contribution} $^{1,3}$\quad \\ \quad \textbf{Gautier Viaud}$^{1}$ \quad \textbf{Antoine Bosselut}$^{4}$ \quad\textbf{Céline Hudelot}$^{3}$\quad 
\textbf{Pierre Colombo}$^{2,3}$ \\ $^{1}$Illuin Technology \quad $^2$Equall.ai \\
$^3$CentraleSupélec, Paris-Saclay \quad $^4$EPFL Lausanne \,\
\\
\small \url{manuel.faysse@centralesupelec.fr}}

\definecolor{maxcomment}{RGB}{111, 8, 201}
\definecolor{navyblue}{RGB}{27, 18, 153}
\definecolor{darkgreen}{RGB}{24, 156, 6}

\newcommand*\colourcheck[1]{%
  \expandafter\newcommand\csname #1check\endcsname{\textcolor{#1}{\ding{52}}}%
}
\colourcheck{blue}
\colourcheck{darkgreen}

\begin{document}

\maketitle
\begin{abstract}
A limitation of modern document retrieval embedding methods is that they typically encode passages (chunks) from the same documents independently, often overlooking crucial contextual information from the rest of the document that could greatly improve individual chunk representations.

In this work, we introduce \emph{ConTEB} (Context-aware Text Embedding Benchmark), a benchmark designed to evaluate retrieval models on their ability to leverage document-wide context. Our results show that state-of-the-art embedding models struggle in retrieval scenarios where context is required. To address this limitation, we propose \emph{InSeNT} (In-sequence Negative Training), a novel contrastive post-training approach which combined with \textit{late chunking} pooling enhances contextual representation learning while preserving computational efficiency. Our method significantly improves retrieval quality on \emph{ConTEB} without sacrificing base model performance. 
We further find chunks embedded with our method are more robust to suboptimal chunking strategies and larger retrieval corpus sizes.
We open-source all artifacts at \url{https://github.com/illuin-tech/contextual-embeddings}. 


\end{abstract}

\section{Introduction}

The ability to rapidly process and query large-scale textual corpora is a cornerstone of many industrial applications, ranging from the analysis of medical records and legal briefs to large-scale administrative archives. As these collections grow in size and complexity, advanced approaches to information retrieval (IR) —particularly Retrieval-Augmented Generation (RAG) \citep{lewis_retrieval-augmented_2020}— have attracted widespread interest, yet, dealing with long documents remains an open challenge.


\begin{figure}[t]
    \centering
    \includegraphics[width=0.35\textwidth]{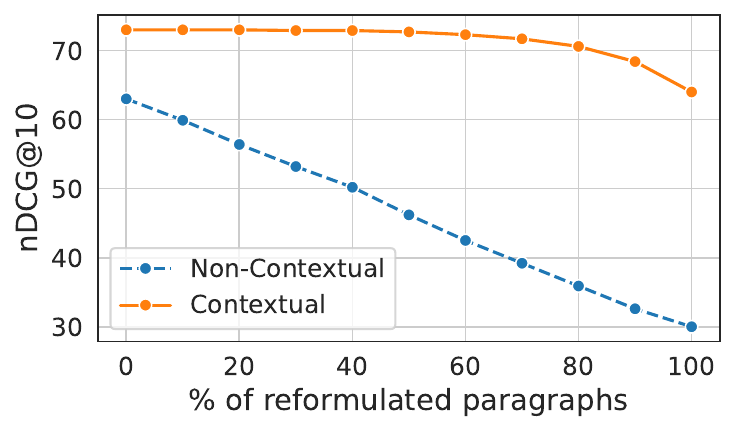}
    \caption{\textbf{Importance of Contextual Information:} Starting from a set of queries and mostly self-contained document paragraphs from the \emph{Football}, we progressively reformulate paragraphs to remove information redundant with the rest of the document. This leads to sharp performance declines in standard retrieval approaches, but not in contextual retrieval approaches.\vspace{-.3cm}}
    \label{fig:reformulated}
    \vspace{-.3cm}
\end{figure}


\begin{figure*}[ht]
    \centering
    \includegraphics[width =\linewidth]{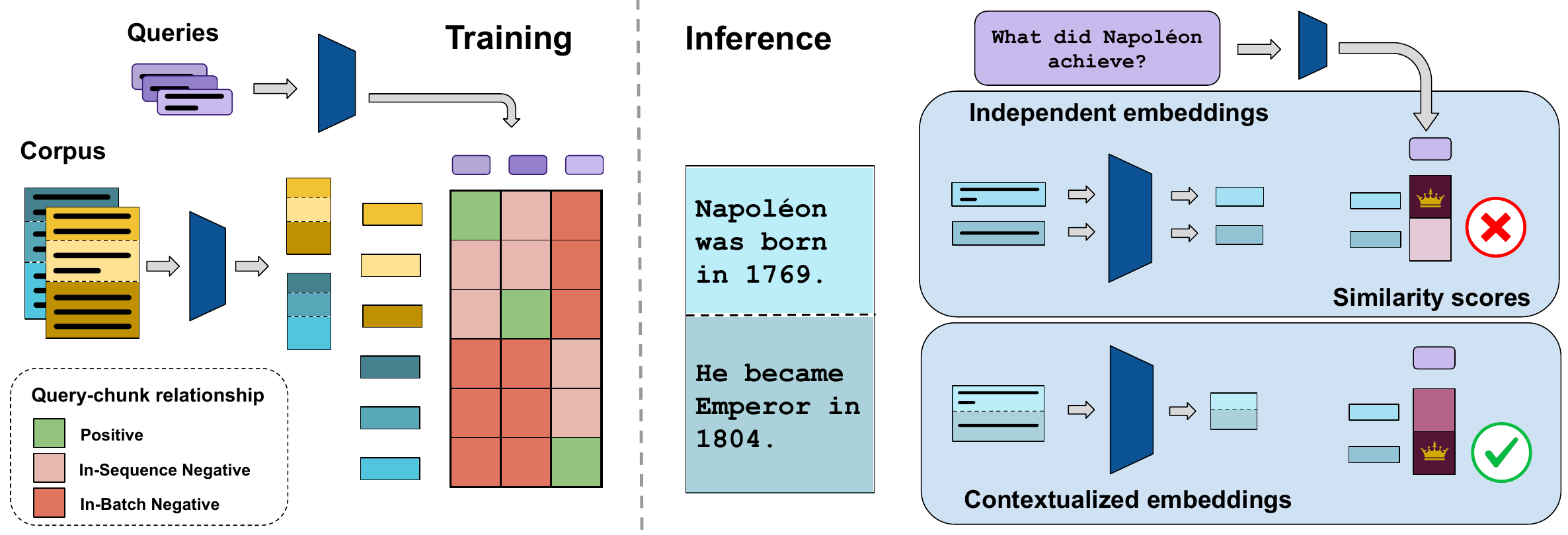}
    \caption{
    \textit{Training (Left).} With respect to a single query, each chunk inside a batch plays a different role, depending on its original document, and the positive chunk.
    \textit{Inference (Right).} Traditional embedding methods (top) produce embeddings that do not include potentially essential contextual information. Contextualized embeddings (bottom) can integrate document-wide information in individual chunk representations, augmenting embedding relevance and improving downstream retrieval performance. \vspace{-.3cm}}
    \label{fig:contextualized_exemple}
\end{figure*}

While long context encoders have been recently developed \citep{zhang2024mgte, modernbert, boizard2025eurobertscalingmultilingualencoders} along with long context embedding models \citep{zhu2024longembedextendingembeddingmodels}, modern document retrieval pipelines typically segment lengthy documents into smaller chunks to optimize the granularity for efficient retrieval and readability of the retrieved content \citep{xu2024retrievalmeetslongcontext, jiang2024longragenhancingretrievalaugmentedgeneration}. Traditionally, these chunks are then \emph{independently} fed to an embedding model, and stored in a vector database for efficient future query matching. By doing so, these systems remove strong semantic and conceptual links between the split passages, directly affecting the resulting representations. An example is illustrated in \autoref{fig:contextualized_exemple}: embedding the sentence \textit{"He became emperor in 1804."} without leveraging information about the person at hand (\emph{Napoléon}) given in previous paragraphs will make matching queries related to \emph{Napoléon} difficult.

Recognizing the significant business value of incorporating broader contextual information into retrieval, major companies have explored leveraging large generative language models (LLMs) to mitigate this limitation. Some approaches attempt to circumvent retrieval altogether by feeding millions of tokens into the model's context window at runtime \citep{geminiteam2024gemini15unlockingmultimodal}, while others reformulate individual passages by concatenating them with document-level summaries and context \citep{contextual_anthropic}. However, these methods are prohibitively expensive at scale when dealing with corpora comprising thousands of documents.

Despite the critical importance of contextualized retrieval, standard benchmarks fail to capture this challenge. Evaluations traditionally focus on assessing the effectiveness of embedding models \citep{thakur_beir_2021, muennighoff_mteb_2022, saadfalcon2024benchmarkingbuildinglongcontextretrieval}, but they rely on datasets where document chunks are by design self-contained answer to the queries, which is a largely idealized scenario in practice \citep{thakur2025freshstackbuildingrealisticbenchmarks}. Consequently, benchmarks fail to highlight the limitations of current retrieval strategies in handling context-dependent passages. Worse, recent findings by \citet{zhou2025gsminfinitellmsbehaveinfinitely} indicate that some widely-used benchmarks exhibit biases that favor standard context-agnostic retrieval methods. Companies such as Anthropic have acknowledged these issues and maintain proprietary contextual retrieval benchmarks that remain unavailable to the public\footnote{\url{https://www.anthropic.com/news/contextual-retrieval}}, underscoring the gap between academic evaluations and real-world industrial needs.


%



\noindent\textbf{Contribution 1: \textit{ConTEB}.}  
We introduce the \emph{Context-aware Text Embedding Benchmark}, designed to assess the ability of retrieval systems to leverage information from the entire document when indexing and retrieving document chunks. \emph{ConTEB} comprises both custom-designed tasks for fine-grained analysis, and practical retrieval evaluation settings spanning multiple document types, domains, and situations in which leveraging context is helpful to produce more meaningful chunk representations. 
We evaluate standard embedding methods on the benchmark and find they struggle when contextual awareness is required.

\noindent\textbf{Contribution 2: \textit{Efficient Contextual Training}.}  
Improving upon the \textit{Late Chunking} method \cite{LateChunking}, we propose a novel embedding post-training method that optimizes information propagation between same-document chunks at indexing time to ensure embeddings are better contextualized. Our method largely boosts performance on \emph{ConTEB}, with minimal computational overhead. 
Through extensive ablations, we detail critical design choices and show our method improves displays increased robustness to sub-optimal chunking strategies and produces representations that scale better with corpus size. 


We open-source all project artifacts, including the benchmark, models and training data\footnote{ \url{https://github.com/illuin-tech/contextual-embeddings}}.

\section{Problem Formulation \& Related Work}

\subsection{Retrieval Frameworks}
In this paper, we consider the traditional retrieval framework where a retrieval system given a query \(q\), searches a corpus \(\mathcal{D}\) for relevant documents. Each document \(d \in \mathcal{D}\) is scored based on its content by first embedding the text into a vector space, and then computing a similarity measure. The similarity between a query \(q\) and a document \(d\) is defined as
\begin{equation}
\text{sim}(q, d) = f\big(\phi(q), \phi(d)\big) \nonumber
\end{equation}
where \(\phi\) maps text into an \(n\)-dimensional vector space and \(f: \mathbb{R}^n \times \mathbb{R}^n \to \mathbb{R}\) is a similarity function, such as cosine similarity or dot product.

In applied settings, individual documents are often too long to be practical for retrieval purposes \citep{Liu_LlamaIndex_2022, zhong2025mixofgranularityoptimizechunkinggranularity}. 
Each document \(d\) is thus divided into segments called \emph{chunks} by a partitioning function \(\mathcal{P}\) defined as
\begin{equation}
\mathcal{P}(d) = \{c_1, c_2, \dots, c_{N_d}\} \nonumber
\end{equation}


In the \emph{standard retrieval} setting, the score is computed solely based on chunk content:

\begin{equation}
\text{sim}(q, c) = f\big(\phi(q), \phi(c)\big) \nonumber
\end{equation}

 Additional information (priors) is however often available to the document embedding system. Typically, knowledge of the entire corpus $\mathcal{D}$, or of \emph{structural metadata} \(M_c\) such as neighboring document chunks obtained through \(\mathcal{P}\), can be leveraged by a modified embedding function \(\phi_2\), yielding the following similarity score:
\begin{equation}
\text{sim}(q, c) = f\Big(\phi(q), \phi_2\big(c, M_c, \mathcal{D}\big)\Big) \nonumber
\end{equation} 

\noindent\textit{This work is centered on efficiently integrating priors about the entire document when embedding a sub-document chunk.}

\subsection{Integrating Contextual Information} \label{related_work_contextual}
Neural embedding models for passage-level text representation, popularized by SentenceBERT \citep{reimers_sentence-bert_2019}, have enabled retrieval systems to move beyond lexical matching \citep{robertson_okapi_1994}. To include contextual information in these retrievers, previous works proposed methods that either operate \textit{offline} during \emph{indexing}, or online during \emph{querying} when faced with a user request.

\noindent\textbf{Indexing.} The chunking strategy is a crucial design choice and often aims to optimize chunk self-containment. Fixed-size approaches with overlaps preserve continuity, while structure-aware chunking respects natural text boundaries, such as paragraphs or sentences. Semantic chunking, by contrast, splits text into topic-aligned segments. These methods appear in frameworks such as LlamaIndex \citep{Liu_LlamaIndex_2022} and LangChain \citep{chase_langchain_2022}, but different queries may need different chunk sizes. Thus, dynamic chunking techniques have emerged to adapt segmentation on the fly \citep{mixofgranularity, qian2024groundinglanguagemodelchunkingfree}. Beyond optimizing chunking, some indexing approaches enrich chunks with broader context by preprending LLM-generated document summaries, contextual information or metadata \citet{contextual_anthropic, multimetarag}. Similarly, \citet{morris2024contextualdocumentembeddings} demonstrate that appending learned "corpus" embeddings to queries and documents can further improve retrieval. Other indexing-time techniques involve organizing chunks into higher-level data structures. For example, \citet{graphrag} and \citet{raptor} cluster related chunks into semantic graphs or tree hierarchies.

\noindent\textbf{Querying.} In contrast, \emph{query-time} solutions rely on iterative or agentic loops to refine retrieval dynamically. LLMs can be used to iteratively update the query or request additional chunks based on partial results \citep{MultiHopRAG, IrCoTRAG}, or even to run “self-checks” and seek extra context when needed \citep{selfrag}. While these adaptive techniques can better address complex, multi-hop queries, they typically require much more computational resources during inference.

\section{\textit{ConTEB}: Context-aware Text Embedding Benchmark}

\label{section:loco_benchmark}

\begin{table*}[ht]
  \centering
  \small

  \sisetup{
    table-number-alignment = center,
    table-figures-integer  = 4,
    table-figures-decimal  = 1,
    table-column-width     = 12mm,
  }

  \definecolor{grpA}{HTML}{C6E0FF}   
  \definecolor{grpB}{HTML}{FFD6D6}   

  \newcolumntype{C}[1]{>{\centering\arraybackslash}p{#1}}

  \renewcommand{\arraystretch}{1.25}

  \begin{tabular}{
      C{5mm}                          
      l                                
      S[table-format=3.1]                
      S[table-format=3.1]                
      S[table-format=3.1]              
      S[table-format=3.1]              
      c
  }
  \toprule
   & Dataset & {Queries} & {Docs} & {\shortstack{Tokens per\\[-1pt]Chunk}} & {\shortstack{Chunks per\\[-1pt]Document}}  & {Context Utilization} \\
  \midrule
  \multirow{3}{*}{{\rotatebox[origin=c]{90}{\shortstack{\textbf{In}\\[-1pt]\textbf{Domain}}}}}
    & MLDR      &  100 &  100 & 170.5 & 15.4 & Document-level reasoning \\
    & NarrativeQA       & 8575 &  355 & 154.5 &  4.9 & Document-level reasoning   \\
    & SQuAD     & 2067 & 2067 &  19.1 &  8.5 & Chunk not self-contained \\
  \midrule
  \multirow{6}{*}{{\rotatebox[origin=c]{90}{\shortstack{\textbf{Out of}\\[-1pt]\textbf{Domain}}}}}
    & {\color{navyblue}\textbf{Football}}   & 2682 & 301 &  77.4 &  20.8 & Co-reference resolution \\
    & {\color{navyblue}\textbf{Geography}}  & 5283 &  530 & 113.6 &  4.3 & Co-reference resolution     \\
    & {\color{navyblue}\textbf{Insurance}}  &  120 &    1 &  80.7 & 60.0 & Structure understanding \\
    & Covid-QA            & 1111 &  115 & 153.9 & 29.1 & Chunk not self-contained  \\
    & ESG Reports         &   36 & 30 & 205.5 & 123.4 & Context disambiguation \\
    & NanoBEIR$^*$        &  650 & 56723 & 199.4 &  1 & No context is needed  \\
  \bottomrule
  \end{tabular}

  \caption{Merged \emph{ConTEB} dataset details. Controlled datasets are highlighted in {\color{navyblue}\textbf{bold blue}}. NanoBEIR values are summed over the 13 datasets that compose it.\vspace{-.3cm}}
  \label{table:merged_conteb}
\end{table*}


\subsection{Benchmark Design}

Existing benchmarks often rely on (or assume) self-contained document chunks. This creates a misleading perception that contextualization offers little to no benefit, which in practice is rarely the case. To address this gap, the \emph{ConTEB} benchmark philosophy is to explicitly be composed of tasks in which leveraging document-wide context should lead to performance improvements. Our benchmark originates from two sources: new datasets specifically created for \emph{ConTEB}, and repurposed academic datasets. We take special care in selecting data sources spanning from multiple domains, including realistic industrial scenarios.

\noindent\textbf{Why Context?} Context can help resolve ambiguity, such as distinguishing between multiple meanings of a word or resolving pronouns and entity references (co-reference resolution). It is also crucial when documents have a structured format, like legal or scientific texts, where understanding table of content hierarchy is key. 

\noindent\textbf{Concept.} To isolate the importance of contextual cues and diminish other confounding factors, we construct three benchmark tasks to study contextualization in controlled experimental settings \citep{AllenZhu-icml2024-tutorial}. 
We also evaluate more practical retrieval settings at larger scale where we suspect contextualization to help, and in which we rely on organic, pre-existing query-document pairs.

\subsection{Benchmark Construction} 
Our generic benchmark curation pipeline is composed of three stages. 
We provide additional curation details in \autoref{app:conteb_stats}. 

\noindent\textbf{1: Chunking.} We select long documents spanning a variety of domains and chunk them through a structure-aware method\footnote{RecursiveCharacterSplitter with a threshold of 1000 characters \citep{chase_langchain_2022}} \citep{rajpurkar-etal-2016-squad, moller-etal-2020-covid, kočiský2017narrativeqareadingcomprehensionchallenge, chen_bge_2024, macé2025vidorebenchmarkv2raising}.

\noindent\textbf{2: Pairing.} 
We use manual answer span annotations (\emph{SQuAD}, \emph{ESG}) or synthetically label them with a LLM (\emph{CovidQA}, \emph{MLDR}, \emph{NarrativeQA}), to match queries with chunks obtained in Stage 1. This ensures queries are not solvable by design \citep{thakur2025freshstackbuildingrealisticbenchmarks}. Alternatively, in our controlled experiment tasks, we generate queries pertaining to the chunks manually (\emph{Insurance}) or synthetically using LLMs (\emph{Football}, \emph{Geography}).

\noindent\textbf{3: Sabotage.} The manually created questions in \emph{Insurance} are designed to be ambiguous without prior knowledge of the document structure. This is manually verified in this phase. Going a step further, in \emph{Football} and \emph{Geography}, we reformulate chunks with the help of a LLM to remove explicit mentions of the original document's theme which all queries mention. We do so in all but the first chunks of each document, explicitly enforcing the need for context.

In addition to our contextual scenarios, we use \emph{NanoBEIR} \citep{thakur_beir_2021} to evaluate non-regression on standard non-contextualized embedding tasks.


\noindent\textit{By combining hard tasks in controlled environments, repurposed academic benchmarks, and real-world industrial queries, our benchmark provides a comprehensive assessment of retrieval models in both standard and context-dependent retrieval scenarios.}

\subsection{Training Dataset}

Open training data is a key factor to ensure fair comparison across methods and robust conclusion. In addition to our benchmark, we construct and release a training dataset composed of query and document chunk pairs. It includes the training splits of \emph{MLDR} and \emph{NarrativeQA}, repurposed with our previously detailed pipeline. To increase the number of queries, we further use \emph{GPT-4o} to generate relevant supplementary synthetic queries. We also concatenate \emph{SQuAD} chunks from the same Wikipedia article, keeping track of the original question-passage associations. 
The full dataset contains 9881 unique long documents (3698 tokens on average), corresponding to a total of 232'587 chunks and 307'241 queries (see \autoref{app:training_data}). Scaling the dataset to more sources, through diverse synthetic augmentations and refinement–based augmentation methods \citep{lee2024nvembedimprovedtechniquestraining, wang2024improvingtextembeddingslarge} is left for future work.

\subsection{Baselines}


\noindent\textbf{Training-Free.} We evaluate a selection of off-the-shelf methods that are strong in their size categories such as a standard single-vector embedding model based on ModernBERT (\texttt{modernbert-embed-large} \citep{warner2024smarterbetterfasterlonger, ModernBERT-embed-large}), its multi-vector ColBERT equivalent \citep{khattab_colbert_2020, GTE-ModernColBERT} and \emph{Okapi BM25} \citep{robertson_okapi_1994}, a strong lexical matching method. 
Additionally, we compare against various contextualization approaches. Specifically, we include Anthropic’s contextual retrieval approach~\cite{contextual_anthropic}\footnote{We use \texttt{Qwen-2.5-7B-Instruct} as the generative model which we serve on a 80GB A100 GPU with vLLM and \texttt{modernbert-embed-large} as the embedding model}, and evaluate Late Chunking \citep{LateChunking} without specific fine-tuning using \texttt{modernbert-embed-large}. These methods cover standard practices with varying level of complexities and indexing budgets.\footnote{We also evaluate RAPTOR ~\citep{raptor} with \texttt{Qwen-2.5-7B-Instruct} and \texttt{cde-small-v2}~\citep{morris2024contextualdocumentembeddings} but find them to be poorly adapted to our problem settings.}

\noindent\textbf{Training-Based.} For fair evaluation, we also fine-tune the sentence embedding method \texttt{modernbert-embed-large} on the training dataset with the same batch construction strategy as when training our main method, ensuring performance differences only stem from methodological design.




\section{Training Contextual Embedders}

In this work, we leverage recent advances in long-context embedding models \citep{zhang2024mgte, modernbert} to improve upon existing approaches through novel training strategies.

\subsection{Architecture}
\label{subsection:architecture}




\noindent\textbf{Late Chunking.}
Late Chunking \cite{LateChunking} (LC) is a training-free token pooling technique designed to enable information propagation across same-document chunks. Formally, given a document $d$ split into chunks $\{c_1, \dots, c_{N_d}\}$, dense retrievers compute independent representations:  
\begin{equation}
\phi(d) = [\phi(c_1), \phi(c_2), \dots, \phi(c_{N_d})] \nonumber
\end{equation}  

In Late Chunking, chunks are concatenated and the whole sequence representation is computed in a single-forward pass:  
\begin{equation}
H = \phi(c_1 \oplus c_2 \oplus \dots \oplus c_{N_d}) \nonumber
\end{equation}  
where $H = [h_1, h_2, \dots, h_T]$ consists of token-level representations.  We then apply average pooling within each original chunk to obtain chunk-wise representations:  
\begin{equation}
\phi_{LC}(c_i) = \frac{1}{|c_i|} \sum_{t \in c_i} h_t, \quad \forall i \in \{1, \dots, N_d\} \nonumber
\end{equation}  

This allows each chunk representation to benefit from contextualization over the full document before aggregation.

\noindent\textbf{Late Interaction.} 
Late Interaction (LI) models \citep{khattab_colbert_2020, chen_bge_2024} are retrieval methods that do not pool token representations and instead store all token embeddings of each document. This approach boosts performance, especially on long-context retrieval tasks \citep{modernbert, zhu2024longembedextendingembeddingmodels}, at the expense of storage cost. In this work, we propose extending Late Chunking approaches to LI models by applying standard LC but simply forgoing the final pooling and storing token embeddings depending on their original chunk memberships.
\begin{equation}
\phi_{LI}(c_i) \;=\; \{\,h_t : t \in c_i\}, 
\quad \forall\,i \in \{1, \dots, N_d\}
\nonumber
\end{equation}

\noindent\textbf{Setup.}
As the base single-vector embedding model for our experiments, we use \texttt{modernbert-embed-large} \citep{ModernBERT-embed-large} (396M parameters), which is fine-tuned for retrieval tasks using the method from \citet{nussbaum2024nomic}.
Respectively, we leverage \texttt{GTE-ModernColBERT} \citep{GTE-ModernColBERT} (149M parameters) for our late interaction experiments. Both models are based on ModernBERT \citep{modernbert} which supports a context length of up to 8,192 tokens, significantly surpassing the 512-token limit of traditional BERT models, and thereby enabling the processing of longer documents in a memory efficient manner, which is critical to our method.



\subsection{Learning Objective}


Late Chunking enables information "leakage" between chunks of the same document. While this training-free method showed promises, we construct a learning objective to explicitly optimize contextual embedding models for this setting.
Our aim is twofold: optimizing chunk representations to integrate relevant document-level information, all while ensuring they retain their specificity with respect to other same-document chunks, in order to prevent embedding collapse.

Previous works \cite{karpukhin_dense_2020, ni2021largedualencodersgeneralizable, izacard_unsupervised_2021, li2023generaltextembeddingsmultistage, wang_text_2022, nussbaum2025nomicembedtrainingreproducible} have relied on various learning objectives inspired by the contrastive learning literature \citep{schroff_facenet_2015}. A natural choice is the InfoNCE objective \citep{oord_representation_2018}, which samples "negative" embeddings from other documents of the same batch. 

In our approach, we combine it with an auxiliary \textit{in-sequence} contrastive loss, where chunks originating from the same document as the positive serve as hard negatives during training. 
Intuitively, training Late Chunking models contrastively with chunks from \emph{different} documents encourages information propagation within each document and improves document identification. On the other hand, the contrastive term between same-document chunks ensures each chunk retains its specificity, and remains identifiable w.r.t. to its neighbors. This aspect is further motivated by the fact that in practice, queried corpora often contain negative documents stemming from the same source. \autoref{fig:contextualized_exemple} illustrates chunk roles across a training batch.


\noindent\textbf{Training Loss.} To balance the contribution of in-sequence and in-batch negatives, we define the weighted InfoNCE loss as:

\begin{equation}\label{eq:loss_weighted}
    \mathcal{L} = \lambda_{\text{seq}} \mathcal{L}_{\text{seq}} + (1 - \lambda_{\text{seq}}) \mathcal{L}_{\text{batch}}
\end{equation}

where \( \lambda_{\text{seq}} \in [0,1] \). Loss terms are defined as:

\begin{equation}
    \mathcal{L}_{\text{seq}} = - \mathbb{E}\left[ \log \frac{\exp\left( q \cdot k^+ / \tau \right)}{\sum_{k_i \in \mathcal{N}_{\text{seq}}} \exp\left( q \cdot k_i / \tau \right)} \right]\nonumber
\end{equation}

\begin{equation}
    \mathcal{L}_{\text{batch}} = - \mathbb{E}\left[ \log \frac{\exp\left( q \cdot k^+ / \tau \right)}{\sum_{k_j \in \mathcal{N}_{\text{batch}} \cup \{k^+\}} \exp\left( q \cdot k_j / \tau \right)} \right]\nonumber
\end{equation}

Here, \( q \) denotes the query representation, and \( k^+ \) is the gold chunk representation, which belongs to \( \mathcal{N}_{\text{seq}} \), the set of chunks from the same sequence as \( k^+ \). Temperature \( \tau > 0 \), and \( \mathcal{N}_{\text{batch}} \) is the set of all in-batch samples that do not belong to \( \mathcal{N}_{\text{seq}} \). This extends to late interaction models by replacing the dot product between query and chunk embeddings by ColBERT's \emph{MaxSim} between the multiple query and document token embeddings.

By tuning \( \lambda_{\text{seq}} \), we can adjust the relative importance of in-sequence versus in-batch contrastive learning (\autoref{fig:lambda_seq}) resulting in our \emph{InSeNT} method.

\subsection{Model training}

Our training strategy (\emph{InSeNT}) is designed to be lightweight and to occur on top of capable pre-trained embedding models without degrading their capabilities. 
We use AdamW, a cosine decay learning rate scheduler with a 5\% warm-up phase and a learning rate of $5e-5$ and train for 2 epochs on our training dataset.
Batches are constructed by sampling 4 long documents per device, retrieving all corresponding chunks and concatenating them with a separator token in between. As documents in our training set contain more than 20 chunks on average, which are themselves often linked to one or multiple queries, a batch contains more than 100 query, positive, negatives triplets to learn on.\footnote{In \textit{MB+Training}, data is sampled the same way for fair evaluation but flattened in batch, corresponding to per-device batch sizes of more than 100.} A single epoch takes less than 1 H100 GPU hour.


\section{Results}
\label{section:results}

\begin{table*}[ht]
  \centering
  \renewcommand{\arraystretch}{1.2}
  \small
  \resizebox{\textwidth}{!}{%
  \begin{tabular}{@{} >{}l ccc cc ccc >{\columncolor{avgcol}}c c c @{} }
    
    & \multicolumn{3}{>{\columncolor{grpA}}c}{\textbf{In-Domain}}  
    & \multicolumn{5}{>{\columncolor{grpB}}c}{\textbf{Out-Of-Domain}}  \\
    \cmidrule(lr){2-9}
    
    & \multicolumn{5}{>{\columncolor{grpD}}c}{\textbf{Practical Settings}}  
    & \multicolumn{3}{>{\columncolor{grpC}}c}{\textbf{Controlled Settings}}  
    & \multicolumn{1}{c}{}  
    & \multicolumn{1}{c}{}  
    & \multicolumn{1}{>{\columncolor{grpD}}c}{\textbf{Non-Contextual}}  \\
    \cmidrule(lr){2-6}\cmidrule(lr){7-9}\cmidrule(lr){12-12}
    
    & \rot{MLDR} & \rot{SQuAD} & \rot{NarrativeQA}
    & \rot{COVID-QA} & \rot{ESG Reports}
    & \rot{Football} & \rot{Geography} & \rot{Insurance}
    & \rot{Average} & \rot{\makecell{Runtime\\(ms/doc)}} & \rot{NanoBEIR} \\
    \midrule
    \multicolumn{12}{@{}l}{\itshape Non‐Contextual Models}\\
    BM25                             & 69.4 & 56.2 & 74.7 & 53.7 & 19.9 & 12.2 & 45.6 & 0.0
                                     & 41.5 & \textbf{4.29}  & 43.4 \\
    ModernBERT Large                 & 78.4 & 73.4 & 77.9 & 61.7 & 36.8 & 19.1 & 56.2 & 12.4
                                     & 52.0 & 17.83 & 63.2 \\
    ModernColBERT                    & 83.5 & 74.2 & 80.4 & \textbf{78.2} & 44.2 & 30.2 & 68.5 & 16.1
                                     & 59.4 & 14.99 & \textbf{67.7} \\
    ModernBERT Large + Training      & 78.7 & 74.0 & 77.3 & 55.2 & 20.0 & 22.9 & 58.7 & 13.9
                                     & 50.1 & 16.44 & 54.5 \\
    \midrule
    \multicolumn{12}{@{}l}{\itshape Untrained Contextual Models}\\
    Anthropic Contextual             & 85.4 & 77.1 & 77.7 & 60.7 & 34.8 & 53.9 & 89.4 & \textbf{100.0}
                                     & 72.4 & 1890.94 & 63.2 \\
    ModernBERT Large + Late Chunking & 78.5 & 77.1 & 75.8 & 40.0 & 31.7 & 54.6 & 89.6 & 41.0
                                     & 61.0 & 15.81   & 63.2 \\
    ModernColBERT + Late Chunking    & 84.1 & 75.7 & 80.7 & 75.5 & 44.4 & 31.3 & 67.9 & 13.2
                                     & 59.1 & 7.41    & \textbf{67.7} \\
    \midrule
    \multicolumn{12}{@{}l}{\itshape Trained Contextual Models}\\
    ModernBERT Large + \emph{InSeNT}          & 88.7 & \textbf{80.9} & 81.3 & 56.0 & 43.1 & 63.9 & \textbf{90.7} & \textbf{100.0}
                                     & \textbf{75.6} & 15.26   & 60.4 \\
    ModernColBERT + \emph{InSeNT}            & \textbf{90.1} & 75.1 & \textbf{83.5} & 67.7 & \textbf{48.3} & \textbf{64.6} & 89.8 & 45.9
                                     & 70.6 & 7.57    & 59.2 \\
    \bottomrule
  \end{tabular}%
  }
  \caption{Evaluation (nDCG@10) of baseline models and our proposed method on \textit{ConTEB}. Runtime is per‐document indexing time in milliseconds; smaller is better, so the fastest model is bolded.}
  \label{table:results}
\end{table*}

\noindent\textbf{Document-wide context is essential.} As seen in \autoref{table:results}, methods leveraging contextual information widely outperform non-contextual methods across \emph{ConTEB} tasks. These results highlight the critical role of context-aware embeddings in improving retrieval performance in such settings, whether through untrained late chunking approaches or expensive context-aware reformulation approaches.
As expected, the gap is even more notable in \emph{ConTEB}'s controlled setting experiments.



\noindent\textbf{Improving contextual information propagation.} Our results clearly show that \emph{InSeNT} variants outperform their untrained counterpart (+14.6 nDCG@10 for ModernBERT, +11.5 for ModernColBERT). Importantly, this is not due to the nature of the training data itself; the non-contextual ModernBERT model trained on the same data (ModernBERT + Training) does not improve upon the untrained baseline. Furthermore, the tasks that display the biggest improvements are the controlled setting tasks \textit{Insurance, Football}, that are explicitly designed to elicit information given in previous paragraphs, and that are out-of-domain w.r.t. our training set.

\noindent\textbf{Late Interaction.} Interestingly, while LI models are good at long-context retrieving, they are poorly suited to out-of-the-box late chunking (-0.3 nDCG@10 w.r.t. ModernColBERT without LI). We posit that since token embeddings are never pooled, these models learn very local features and cannot leverage information from neighboring tokens. Once trained with our method, ModernColBERT+\emph{InSeNT} displays large performance gains across the board (+11.5 nDCG@10 w.r.t. ModernColBERT + Late Chunking), showcasing an increased ability to leverage external context.

\noindent\textbf{Context can add noise.} The \emph{CovidQA} task sticks out from the rest as untrained late chunking approaches severely degrade performance. Qualitative analysis, as well as the strong performance of the non-contextualized ModernColBERT method, indicate that the query-chunk pairing are often very extractive and match on technical medical terms, thus rendering context less useful. Our results show that naively applying late chunking in this setting adds noise and leads to notable performance drops (-21 nDCG@10), which are in large part recovered through our training method (+16 nDCG@10). 



\begin{figure}[btp!]
    \centering
    \includegraphics[width=0.45\textwidth]{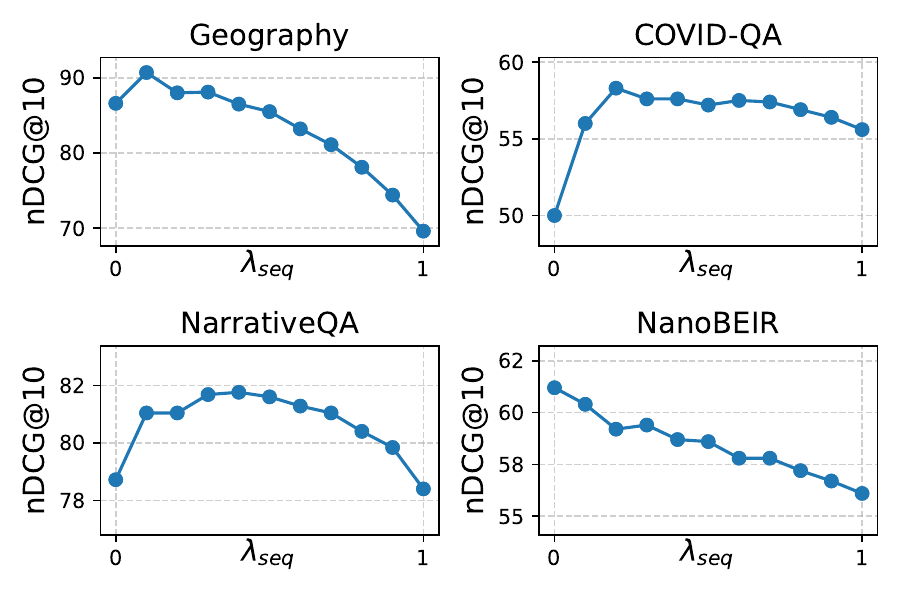}
    \caption{\textbf{\textbf{Importance of $\lambda_{seq}$}:} Results for ModernBERT-Large trained with varying $\lambda_{seq}$. Optimal values depend on the task, but integrating both in-sequence and in-batch negatives is crucial to performance.\vspace{-.6cm}}
    \label{fig:lambda_seq}
\end{figure}
\label{ablation_lambda_seq}

\noindent\textbf{$\lambda_{seq}$ matters.} The training objectives are to induce chunk representations to integrate document-level information (role of \emph{in-batch} negatives) while maintaining their specificity with respect to other same document chunks (role of \emph{in-sequence} negatives). By varying $\lambda_{seq}$ from \autoref{eq:loss_weighted}, we weight the importance of both objectives.

\noindent After training a series of models with varying $\lambda_{seq}$, we see on \autoref{fig:lambda_seq} that training with only in-sequence or in-batch negatives yields the worse results, and the optimal $\lambda_{seq}$ varies depending on the task. When documents need to be disambiguated between one another (\emph{NanoBEIR}, \emph{Geography}), up-weighting in-batch negatives seems optimal. On tasks where the challenge lies in locating information within a given document (\emph{NarrativeQA}, \emph{Covid-QA}), in-sequence negatives play a large role, but still need to be combined to in-batch negatives. Striking the optimal trade-off is thus very use-case dependent, and we opt for $\lambda_{seq}=0.1$ after tuning on the validation split of our training dataset.


\begin{figure*}
    \centering
    \includegraphics[width = 1\textwidth]{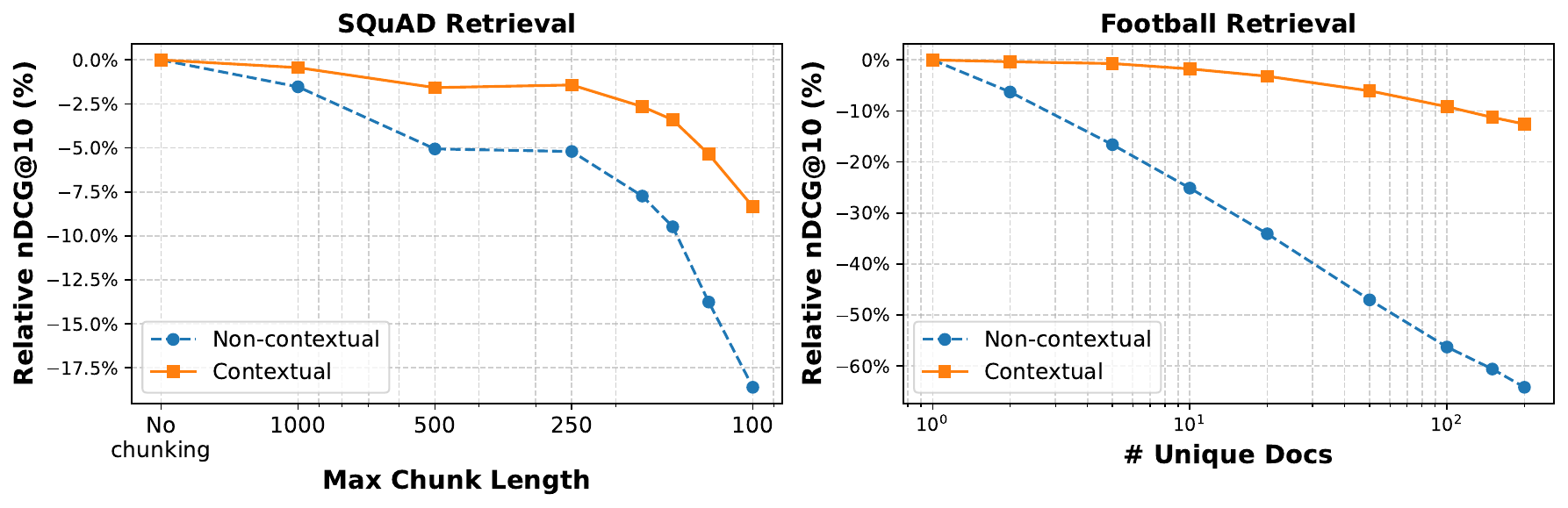}
    \caption{Contextualized models trained with InSeNT are more robust to aggressive chunking strategies that remove essential information from chunks (left), and scale better with corpus size and ambiguity (right).}
    \label{fig:robustness}
\end{figure*}

\noindent\textbf{Efficiency-Performance.} As shown in the Runtime column of \autoref{table:results}, our approach is very capable on contextual tasks, yet does not add much computational overhead. In fact, we find slight indexing speed improvements, attributed to our approach's reduced need for padding in-batch sequences of different lengths. While \textit{Anthropic Contextual} achieves sensibly similar performances on \emph{ConTEB}, it relies on costly LLM-based summarization and chunk reformulation, that are hardly scalable to huge corpora (120x slower).

\noindent\textbf{Short-Context Performance.} Careful hyperparameter tuning enables our best model to maintain strong performance on standard non-contextual benchmarks (\textit{NanoBEIR}), demonstrating that long-context optimization does not compromise short-context retrieval. Interestingly, LI models suffer from more degradation, which we posit is due to the original reliance on very local features modified through our training. Mixing in non-contextual "replay" data during training or merging models \cite{modelmerging_lines} should further enable preserving the original embedding model's performances.


\section{Ablations}







\noindent\textbf{Robustness to chunking.}
We assess our method's robustness to poor chunking strategies using \emph{SQuAD} annotations.  Each originally self-contained chunk is split in multiple progressively smaller sub-chunks to while we keep track of the annotated answer span to identify the gold chunk. Eventually, these sub-chunks become too small to be self-contained and end up lacking sufficient information to be relevantly embedded on their own. \autoref{fig:robustness} (left) demonstrates that contextual embeddings greatly improves robustness w.r.t. suboptimal chunking. The model is able to elicit information from neighboring chunks to integrate contextual information within smaller sub-chunks, leading to a much more uniform retrieval performance across a wide range of chunk sizes.

\noindent\textbf{Robustness to corpus size.}
Common in the industry are templated documents that differ mostly by a key aspect (year, company name) but contain otherwise very similar information.
We study the dynamics of retrieval performance w.r.t. to the amount of similar documents in the corpus by computing scaling laws in which we iteratively vary the number of unique documents (composed of multiple chunks) in the corpus.
We observe in \autoref{fig:robustness} (right) that contextual embeddings scale vastly differently than their independently embedded counterpart. 
Intuitively, the greater the amount of similar documents and chunks in the corpus, the harder it is for a retrieval system to match the correct ones, but when embedding models are able to leverage external context, this effect is attenuated.

\noindent\textbf{Information Propagation.}
We experiment with concatenating semantically similar yet independent short chunks as "artificial" long documents. The resulting model is contextual as it uses late chunking, but exhibits performances in-line with non-contextual baselines (\textit{ModernBERT Large + Training}). We posit training on arbitrarily concatenated chunks, which by design are not contextually linked, teaches the model not to use information from neighboring chunks. This highlight the necessity of sourcing organic long-context data during training to induce correct training dynamics. Details in \autoref{table:nomic-results} in \autoref{app:res}.






\section{Conclusions}
\label{section:conclusions_and_future_work}

In this work, we introduced \textit{ConTEB}, a benchmark designed to assess the effectiveness of retrieval models in leveraging document-wide contextual information. Our evaluation demonstrates that standard retrieval models struggle in context-dependent settings, while our proposed approach \textit{InSeNT}, which combines late chunking and a novel training methodology performs strongly on \textit{ConTEB} without additional compute costs. 

\noindent\textbf{Future Work.} Scaling our approach with recent decoder models with extended context lengths (e.g., 1M+ tokens \citep{yang2025qwen251mtechnicalreport}) would enable embedding entire books or lengthy documents in a single forward pass, potentially unlocking new capabilities for large-scale document retrieval. It would also be interesting to observe the impact of our method on retrieval confidence \citep{gisserotboukhlef2024trustworthyrerankingsimpleeffective}. Finally, adapting our method to multi-modal embedding pipelines that have less control over the chunking strategy could further enhance retrieval systems in industrial applications with visually rich contextual documents \citep{faysse2025colpaliefficientdocumentretrieval, ma2024unifyingmultimodalretrievaldocument}. 


\section*{Limitations}

While our approach enhances retrieval performance in context-dependent settings, limitations persist.

\noindent\textbf{Context Length.} Our method is applied to long-context encoders that currently support sequences of up to 8k tokens. While we have shown we can extrapolate performance to sequences of up to 32k tokens, scaling this approach to handle 1M+ token contexts with decoder-based models would be an interesting research avenue and presents significant compute and memory challenges. Additionally, it requires rethinking the data construction processes to ensure longer documents are effectively leveraged.

\noindent\textbf{Data Generation.} The creation of training and evaluation data relies on existing datasets and semi-synthetic generation pipelines. However, a fully automated and scalable method for generating high-quality queries that effectively induce non-trivial context utilization remains an open challenge.

\noindent\textbf{Evaluation.} While our model demonstrates strong cross-domain performance, further validation in real-world applications, various use cases, and multiple languages is necessary to further assess its robustness and generalizability.

\section*{Ethical Considerations}

\noindent\textbf{Bias.} As our method introduces a novel way of leveraging document-wide context, the nature of information propagation between chunks remains uncertain. This may introduce biases that traditional embedding models do not encounter, necessitating further analysis.

\noindent\textbf{Ecological Impact.} Our post-training approach is computationally efficient, with total training and evaluation runs requiring fewer than 100 GPU hours on H100 hardware. By providing a cost-effective alternative to LLM-dependent contextualization techniques, we aim to reduce the environmental footprint of large-scale retrieval systems.

\noindent\textbf{Social Impact.} Improved retrieval capabilities can drive significant business benefits, particularly in industries that rely on processing extensive and structured documents, such as legal, medical, and financial sectors.






\bibliography{references}

\appendix
\label{appendix}

\section{\textit{ConTEB} Details}\label{app:conteb_stats}

\begin{figure*}[htbp]
  \centering
  \includegraphics[width=\textwidth]{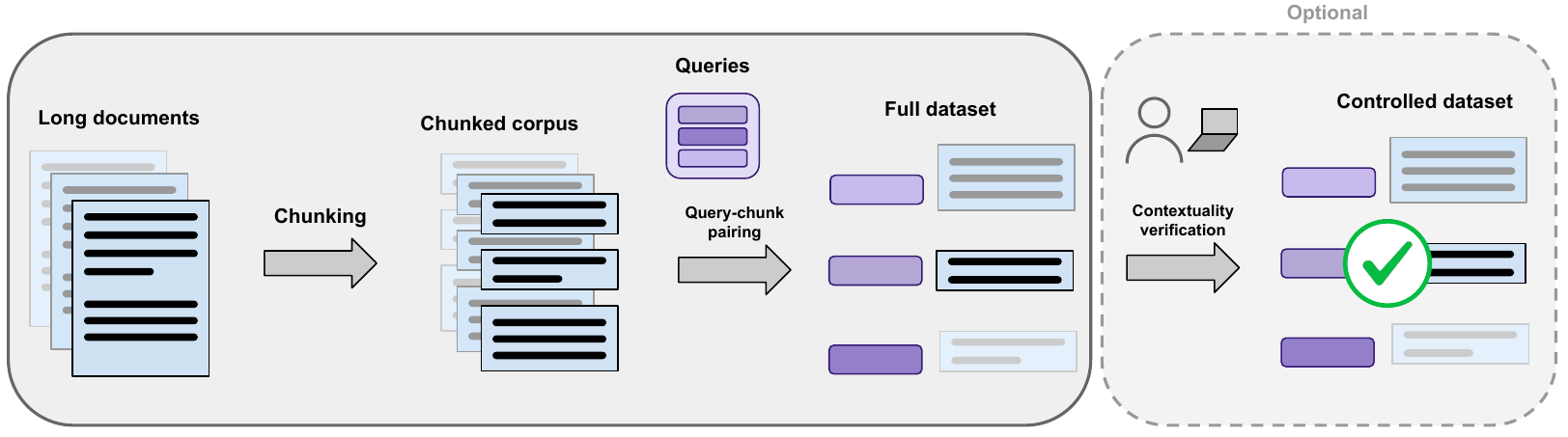}
  \caption{Benchmark creation process.}
  \label{fig:benchmark_creation}
\end{figure*}

This appendix describes the data generation process employed in this project. The methodology varies based on the dataset source, but generally, long documents are segmented into smaller chunks. If preexisting queries are available, they are mapped to relevant chunks using either provided answer spans (e.g., SQuAD) or tagged using GPT-4o. In cases where queries are unavailable, a large language model (LLM) generates them before associating them with the relevant text segments. This approach, illustrated in \ref{fig:benchmark_creation}, is systematically applied across multiple datasets.

\subsection{Wiki-based Datasets} \label{app:wiki-based}

\emph{Football} and Geography are our two wiki-based datasets, focusing on the Sports and Geography domains.

\noindent\textbf{Wikipedia Data Extraction}
The pipeline first retrieves Wikipedia summaries for a given person using the \texttt{wikipediaapi} library. The extracted summary is then split into paragraphs.

\noindent\textbf{Text Rephrasing}
Each paragraph from the Wikipedia summary undergoes a rephrasing process to remove direct mentions of the person’s name while maintaining the original context. The rephrased text replaces names with pronouns such as ‘he’ or ‘she’. This transformation is performed using the GPT-4o model via the following prompt:

\begin{quote}
\textit{
Here is a Wikipedia article:
\texttt{[Full Wikipedia Summary]}
Can you rephrase the following paragraph to remove all mention of the name of the person the article is about? You can leave other names as is and can replace the name with words such as 'he/she' or other generic paraphrases.
\texttt{[Paragraph to be rephrased]}
}
\end{quote}

\noindent\textbf{Question Generation}
For each paragraph in the summary, the model generates three questions related to the person. The questions explicitly mention the person's name but do not include other named entities such as dates or proper nouns. The generation follows this structured prompt:

\begin{quote}
\textit{
Here is a Wikipedia article:}

\texttt{[Full Wikipedia Summary]}

Using specifically the following paragraph, can you ask 3 questions related to the person the article is about? Each question must mention the name of the person, but the question should not contain other named entities (dates, other proper nouns). Format the response as a Python list of strings and do not output anything else.

\texttt{[Paragraph to be used for question generation]}

\end{quote}

\subsection{NarrativeQA, COVID-QA, MLDR}

\textit{NarrativeQA} (literature), \textit{MLDR} (encyclopedic) and \textit{Covid-QA} (medical) consist of long documents, associated to existing sets of question-answer pairs. 

We chunk these documents, and use GPT-4o to annotate which chunk, among the gold document, best contains information needed to answer the query. Since chunking is done \textit{a posteriori} without considering the questions, chunks are not always self-contained and eliciting document-wide context can help build meaningful representations. 

\textbf{Synthetic Query Generation:} To extend MLDR for our training dataset, OpenAI's GPT-4o model is prompted to generate 20-50 realistic queries per document, ensuring that each query aligns with the content of at least one chunk. This is on top of the queries that are already incuded in the dataset. Synthetic queries are included only in our training dataset.

\subsection{Insurance} 
\textit{Insurance} is composed of a long document with insurance-related statistics for each country of the European Union. Countries are often not referred to in-text, but only once in the section title. Therefore, certain chunks require knowledge of their position within the document to be properly disambiguated from others. Questions are manually crafted to require structural understanding for accurate chunk matching. This process, in addition to manual verification of the contextuality quality, makes \textit{Insurance} a controlled dataset. Since questions are crafted after the chunking process, the annotation results directly from the manual question generation process.

\subsection{SQuAD}

\textit{SQuAD} is an extractive QA dataset with questions associated to passages and annotated answer spans, that allow us to chunk individual passages into shorter sequences while preserving the original annotation.

\subsection{ESG Reports}

\textit{ESG Reports} contains long documents from the fast-food industry, with manually annotated query-page pairs from the ViDoRe Benchmark v2 \citep{macé2025vidorebenchmarkv2raising}, originally thought for visual retrieving\footnote{\url{https://huggingface.co/datasets/vidore/restaurant_esg_reports_beir}}. We convert all documents to text, chunk them, and re-annotate the resulting passages by hand, filtering out queries that relied solely on visual aspects (e.g., tables, graphs).


\subsection{Training Data Statistics}\label{app:training_data}

\autoref{table:training-data} displays information about the training data. Our refined version of MLDR forms a large part of the training corpus. We can see that the majority of chunks are used as positives at least once, ensuring that the model is not biased towards the position of the chunk in the sequence.

\begin{table}[ht]
    \centering
    \resizebox{\linewidth}{!}{%
    \begin{tabular}{l|rrr|r}
    \toprule
     & MLDR & NarrativeQA & SQuAD & Total \\
    \midrule
    Number of Docs & 8467 & 972 & 442 & 9881 \\
    Number of Chunks & 213001 & 5219 & 14367 & 232587 \\
    Number of Queries & 211933 & 27953 & 67355 & 307241 \\
    Number of Chunks per Doc & 25.2 & 5.4 & 32.5 & 23.5 \\
    \% Chunks with associated Query & 94.6\% & 81.9\% & 100.0\% & 94.61\% \\
    Number of Tokens per Doc & 3962.6 & 819.1 & 4966.1 & 3698.2 \\
    Number of Tokens per Query & 16.7 & 21.9 & 12.5 & 16.3 \\
    \bottomrule
    
    \end{tabular}
    }
    
    \caption{Training Dataset Statistics} 
    \label{table:training-data}
\end{table}






\section{Implementation Details}

\subsection{Sequence prefixes}

ModernBERT-based models are trained with query and document prefixes. We apply the same approach in our training and inference frameworks. After several tests, we opt for using a single document prefix for the Late Chunking sequence, instead of adding a document prefix at the beginning of each chunk inside the same sequence. We separate chunks with [SEP] tokens to let the model understand the concept of chunks during its token embedding computation.

\subsection{Late Interaction Models}

We leverage the \texttt{pylate} \citep{PyLate} library for the Late Interaction implementation. For training LI models with InSeNT, we adapt the LI mechanisms to incorporate it with Late Chunking in our own codebase. In particular, we do not use token skiplists at inference time, and use a single document prefix for the whole document sequence.

\section{Additional Results}
\label{app:res}

\subsection{Training with concatenated short documents}

Results of training an InSeNT model with concatenated short document data (using the Nomic dataset) are available in \autoref{table:nomic-results}. Short docs are clustered from the \texttt{nomic-supervised} dataset \citep{nussbaum2024nomic} following \citet{morris2024contextualdocumentembeddings}. This approach did not yield promising results, proving that natively long documents are necessary to induce relevant in-sequence signal.

\begin{table*}[ht]
    \centering
    \renewcommand{\arraystretch}{1.2} 
    \resizebox{\textwidth}{!}{%
    \begin{tabular}{l|rrr|rrrr|r|rr}
\toprule
     & MLDR & SQuAD & NarrativeQA & Football & Geography & COVID-QA & Insurance & NanoBEIR & Average & Runtime (s) \\
    \midrule
    MB & 78.4 & 73.4 & 77.9 & 19.1 & 56.2 & \textbf{61.7} & 12.4 & \textbf{63.2} & 55.3 & 40.0\\
    \midrule
    MB+InSeNT(Nomic) & 77.8 & 76.0 & 76.2 & 26.2 & 62.7 & 38.8 & 63.7 & 59.9 & 60.2 & 36.3\\
    MB+Late Chunking & 78.5 & 77.1 & 75.8 & 54.6 & 89.6 & 40.0 & 41.0 & \textbf{63.2} & 65.0 & 36.3 \\
    \textbf{Ours: MB+InSeNT} & \textbf{88.7} & \textbf{80.9} & \textbf{81.3} & \textbf{63.9} & \textbf{90.7} & 56.0 & \textbf{100.0} & 60.4 & \textbf{77.8} & 36.3 \\
    \bottomrule
    \end{tabular}}

    \caption{Evaluation (nDCG@10) of baseline models and our proposed method on \textit{ConTEB}. We show MB+InSeNT(Nomic) behaves like a non-contextual model after training on independant documents concatenated in a single sequence.} 
    \label{table:nomic-results}
\end{table*}

\subsection{Full ablation results on $\lambda_{seq}$}

We show the results of the different values for $\lambda_{seq}$ on all our evaluation sets.



\begin{figure*}
    \centering
    \includegraphics[width=0.49\linewidth]{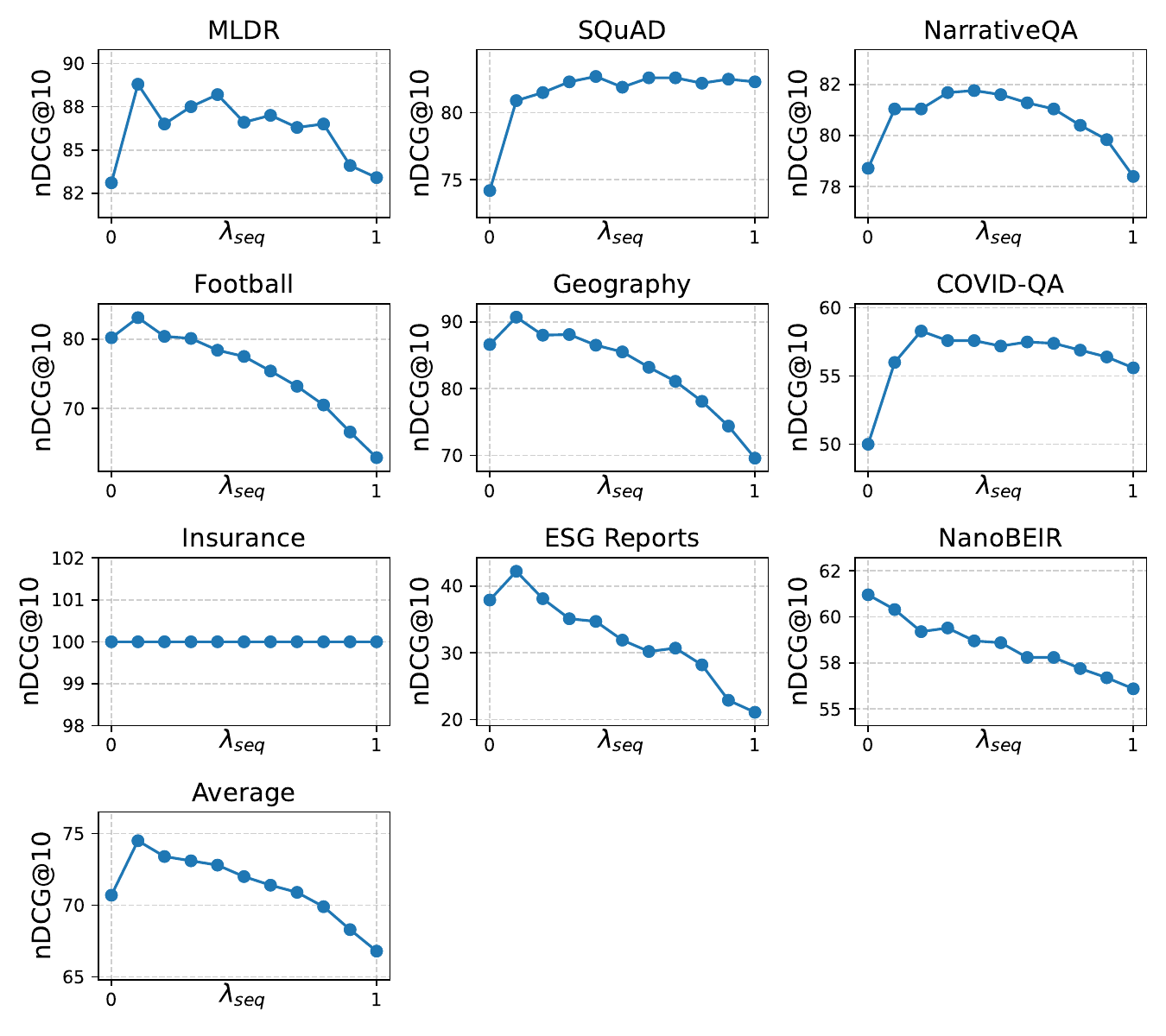}
    \hfill
    \includegraphics[width=0.49\linewidth]{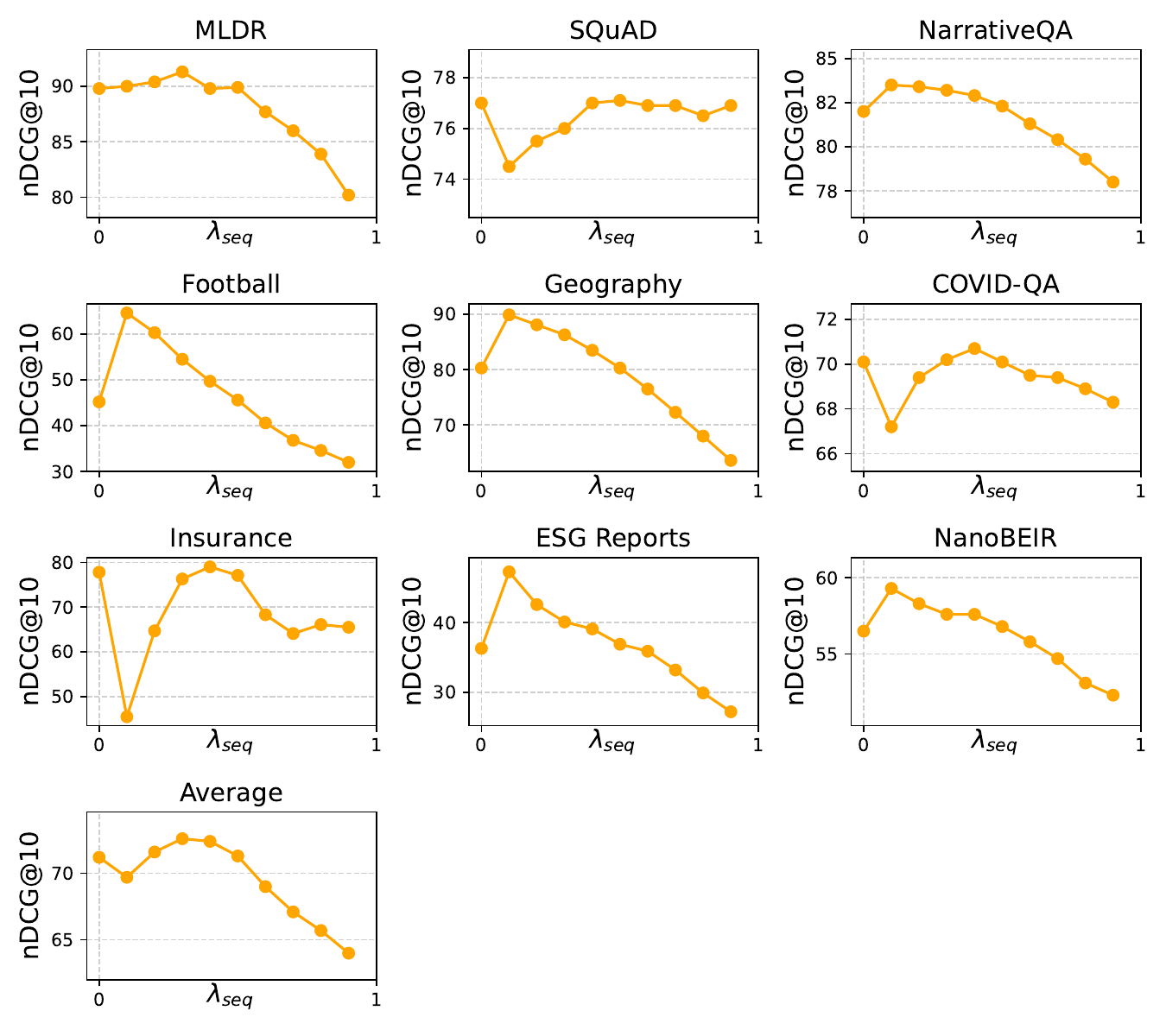}
    \caption{Evaluation results for varying $\lambda_{seq}$ values. Left: ModernBERT-Large. Right: GTE-ModernColBERT. Trends vary across the datasets depending on their nature.}
    \label{fig:lambda_seq_side_by_side}
\end{figure*}

\subsection{Extending context beyond 8192 tokens}

ModernBERT was trained on documents of up to 8192 tokens \cite{modernbert}. Its Late Interaction counterpart, GTE-ModernColBERT, was exclusively fine-tuned on documents of no more than 300 tokens. However, its generalization capabilities to longer documents have been shown by its developers \cite{GTE-ModernColBERT}, hinting at the fact that further research along those lines could be tried for both the bi-encoder and the LI variants.

Based on these results, we tried two approaches to handle documents longer than 8192 tokens with ModernBERT (necessary for the ESG reports dataset): computing Late Chunking with a context of max. 8192 tokens in an sliding window fashion (computing chunk embeddings in several forward passes of 8192 tokens, with 10 overlapping chunks between the various windows), and naively feeding the complete documents to the embedder. 

To our surprise, the latter worked better by a large margin (43.1 on ESG as reported in \ref{table:results}, vs 25.4 for the sliding window approach), so we reported the results of this approach. Further studies could be led to better understand the dynamics underlying this extension.

\end{document}